\documentstyle[prl,aps,multicol,psfig]{revtex}
\begin{document}
\draft
\title{Sharp increase of the effective mass near the critical
density\\ in a metallic 2D electron system}
\author{A.~A. Shashkin$^*$ and S.~V. Kravchenko}
\address{Physics Department, Northeastern University, Boston,
Massachusetts 02115}
\author{V.~T. Dolgopolov}
\address{Institute of Solid State Physics, Chernogolovka, Moscow
District 142432, Russia}
\author{T.~M. Klapwijk}
\address{Department of Applied Physics, Delft University of
Technology, 2628 CJ Delft, The Netherlands}
\maketitle
\begin{abstract}
We find that at intermediate temperatures, the metallic temperature
dependence of the conductivity $\sigma(T)$ of 2D electrons in silicon
is described well by a recent interaction-based theory of Zala {\it
et al.} (Phys.\ Rev.\ B\ {\bf 64}, 214204 (2001)). The tendency of
the slope $\sigma^{-1}d\sigma/dT$ to diverge near the critical
electron density is in agreement with the previously suggested
ferromagnetic instability in this electron system. Unexpectedly, it
is found to originate from the sharp enhancement of the effective
mass, while the effective Land\'e $g$ factor remains nearly constant
and close to its value in bulk silicon.
\end{abstract}
\begin{multicols}{2}

In dilute two-dimensional (2D) electron systems, the energy of
electron-electron interactions dominates the kinetic energy making
the system strongly correlated \cite{rmp}. The interaction strength
is characterized by the Wigner-Seitz radius, $r_s$, which is equal in
the single-valley case to the ratio of the Coulomb and the Fermi
energies. According to the Fermi liquid theory \cite{landau} (whose
applicability to dilute 2D electron systems is discussed in
Ref.~\cite{castro}), the electron-electron interactions should give
rise to a renormalization of the system parameters including the
effective electron mass, $m$, and the effective $g$ factor. A sharp
enhancement of the product $gm$ with decreasing electron density ---
possibly, a precursor of the ferromagnetic instability --- has been
observed in recent studies of the parallel field magnetoresistance of
a metallic 2D electron system in high-mobility silicon
metal-oxide-semiconductor field-effect transistors (MOSFETs)
\cite{ferro}. This agrees well with the $gm$ value obtained from the
beating pattern of Shubnikov-de Haas oscillations in tilted magnetic
fields \cite{krav,pudalov01,comment,PGK}.

Recently, temperature-dependent corrections to conductivity due to
electron-electron interactions have been calculated by Zala {\it et
al.} \cite{aleiner} based on the Fermi liquid approach. In contrast
to pre-existing theories (see, e.g., Refs.~\cite{GD,sarma}), the new
theory incorporates strongly interacting 2D electron systems with
electron densities, $n_s$, down to the vicinity of the
metal-insulator transition (provided that the conductivity $\sigma\gg
e^2/h$). For sufficiently strong interactions, it predicts a metallic
temperature dependence of conductivity in the entire temperature
range. At very low temperatures, in the ``diffusive'' regime
($T\ll\hbar/k_B\tau$, where $\tau$ is the elastic relaxation time),
this is Finkelstein's weakly-metallic (logarithmic) conductivity
\cite{finkelstein83,punnoose01}. At intermediate temperatures, in the
``ballistic'' regime ($T\agt\hbar/k_B\tau$; $T>0.2-0.5$~K under the
conditions of our experiments), the predicted $\sigma(T)$ is similar
to the Gold-Dolgopolov linear dependence \cite{GD}:

\begin{equation}
\frac{\sigma(T)}{\sigma_0}=1-Ak_BT, \label{sigma}\end{equation}
where the slope, $A$, is determined by the interaction-related
parameters: the Fermi liquid constants, $F_0^a$ and $F_1^s$. These
parameters are responsible for the renormalization of the $g$ factor
and the effective mass \cite{castro}

\begin{equation}
\frac{g}{g_0}=\frac{1}{1+F_0^a},\qquad\frac{m}{m_b}=1+F_1^s \label{F}
\end{equation}
and can be determined experimentally \cite{rem0}. The slope $A$ is
predicted to rise as the interaction strength increases and the 2D
electron system is driven toward the ferromagnetic instability. The
last is expected to occur, in the simplest case, at $F_0^a=-1$, which
corresponds to the diverging effective $g$ factor.

In this paper, we perform precision measurements of the
temperature-dependent conductivity in a metallic 2D electron system
in silicon over a wide range of electron densities above the critical
electron density, $n_c$, for the metal-insulator transition. The
theory of Zala {\it et al.} \cite{aleiner} is found to be successful
in interpreting the experimental data in the ballistic regime.
Knowing the product $gm$ from independent measurements, we determine
both $g$ and $m$ as a function of $n_s$ from the slope of the
temperature dependence of the conductivity. The tendency of the slope
to diverge near the critical density is consistent with the suggested
ferromagnetic instability in this electron system \cite{ferro,rem2}.
However, unlike in the simplest scenario for the ferromagnetic
instability, it is the value of the effective mass that becomes
strongly enhanced with
\vbox{
\vspace{-20mm}
\hbox{
\hspace{2mm}
\psfig{file=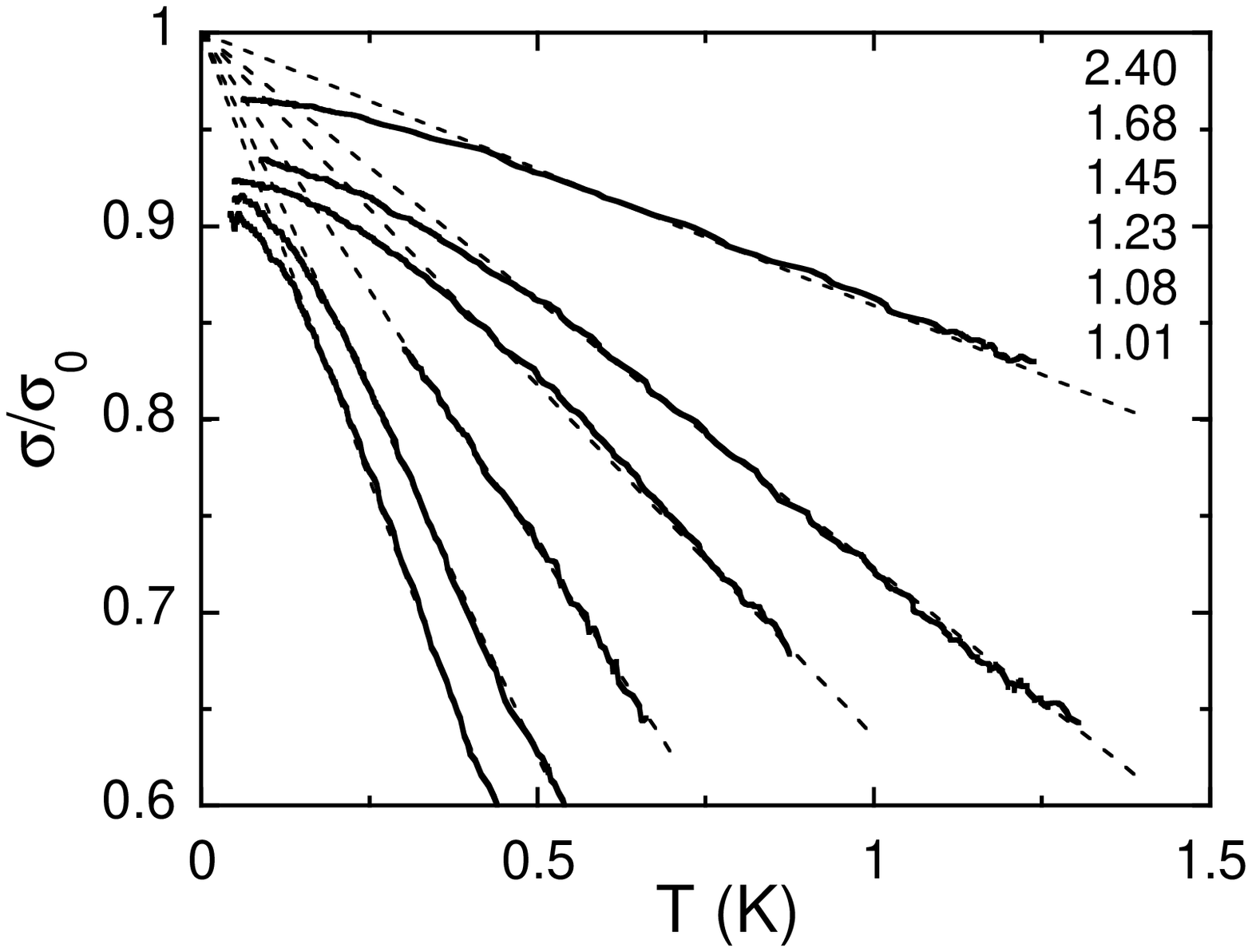,width=2.9in,bbllx=.5in,bblly=1.25in,bburx=7.25in,bbury=9.5in,angle=0}
}
\vspace{-0.2in}
\hbox{
\hspace{-0.15in}
\refstepcounter{figure}
\parbox[b]{3.4in}{\baselineskip=12pt \egtrm FIG.~\thefigure.
The temperature dependence of the normalized conductivity at
different electron densities (indicated in units of
$10^{11}$~cm$^{-2}$) above the critical electron density for the
metal-insulator transition. The dashed lines are fits of the linear
interval of the dependence.\vspace{0.20in}
}
\label{fan}
}
}
decreasing electron density, while the $g$
factor remains nearly constant, $g\approx g_0=2$ in bulk silicon.

Measurements were made in an Oxford dilution refrigerator with a base
temperature of $\approx30$~mK on high-mobility (100)-silicon samples
similar to those previously used in Ref.~\cite{krav00}. The
resistance was measured by a standard 4-terminal low-frequency
technique. Excitation current was kept low enough to ensure that
measurements were taken in the linear regime of response. Contact
resistances in our samples were minimized by using a split-gate
technique that allows one to maintain a high electron density in the
vicinity of the contacts regardless of its value in the main part of
the sample. In this paper we show results obtained on a sample with a
peak mobility close to 3~m$^2$/Vs at 0.1~K.

Typical dependences of the normalized conductivity on temperature,
$\sigma(T)/\sigma_0$, are displayed in Fig.~\ref{fan} at different
electron densities above the critical electron density for the
metal-insulator transition which in this sample occurs at
$n_c=8\times10^{10}$~cm$^{-2}$ \cite{rem}; the value $\sigma_0$,
which has been used to normalize $\sigma$, was obtained by
extrapolating the linear interval of the $\sigma(T)$ dependence to
$T=0$. As long as the deviation $|\sigma/\sigma_0-1|$ is sufficiently
small, the conductivity $\sigma$ increases linearly with decreasing
$T$ in agreement with Eq.~(\ref{sigma}), until it saturates at the
lowest temperatures \cite{rem1}.

In Fig.~\ref{slope}, we show the $n_s$ dependence of the inverse
slope $1/A$ extracted from the $\sigma(T)$ data. Also shown for
comparison is the magnetic energy, $\mu_BB_c=\pi\hbar^2n_s/gm$ (where
$\mu_B$ is the Bohr magneton), corresponding to the onset of full
spin polarization in this electron system, which is governed by the
(enhanced) product $gm$ \cite{ferro}. Over a wide range of electron
densities, the values $1/A$ and $\mu_BB_c$ turn out to be close to
each other. The low
\vbox{
\vspace{-17mm}
\hbox{
\hspace{2mm}
\psfig{file=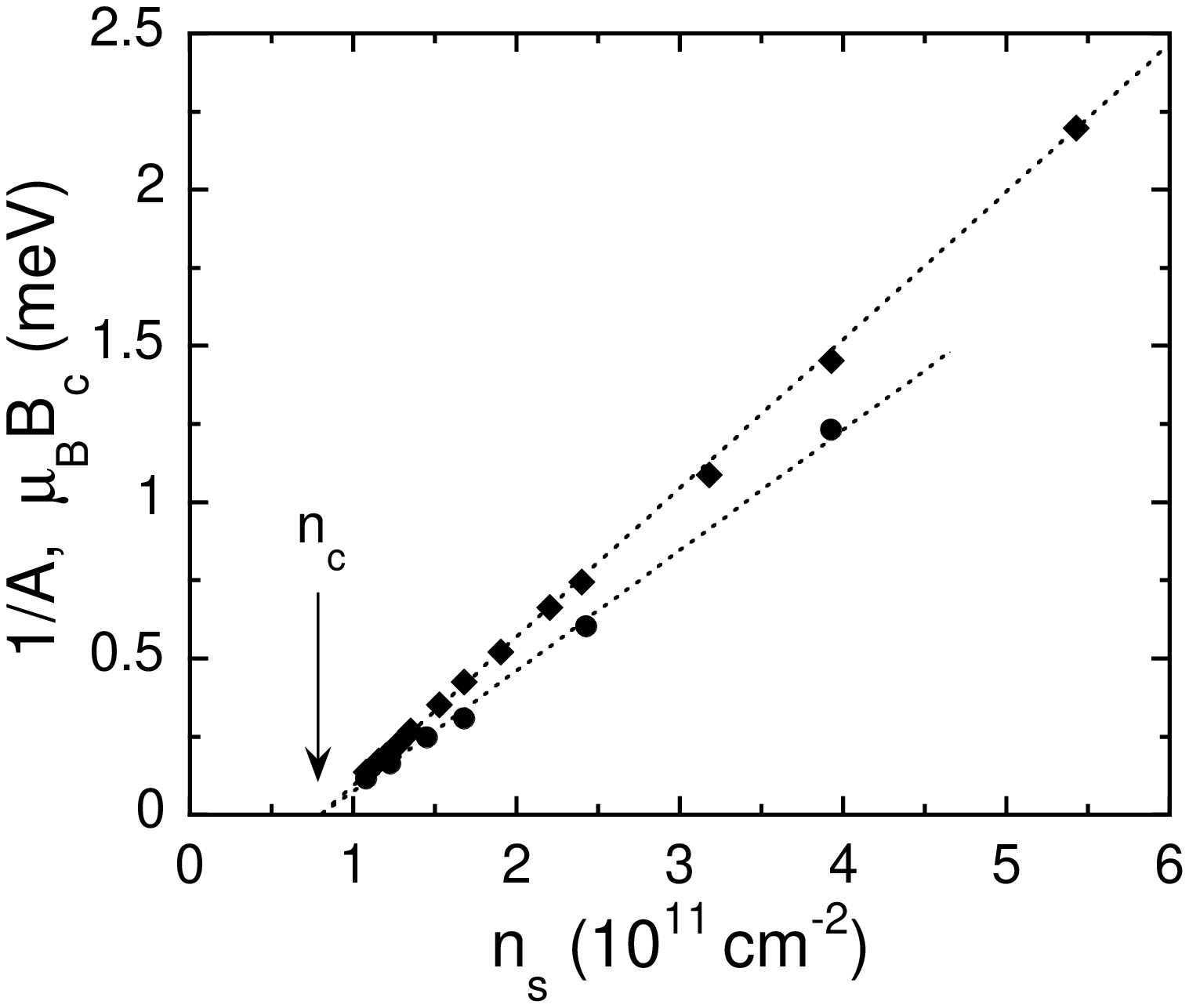,width=2.9in,bbllx=.5in,bblly=1.25in,bburx=7.25in,bbury=9.5in,angle=0}
}
\vspace{-0.3in}
\hbox{
\hspace{-0.15in}
\refstepcounter{figure}
\parbox[b]{3.4in}{\baselineskip=12pt \egtrm FIG.~\thefigure.
Comparison of the inverse slope $1/A$ (dots) and the data for the
polarization field $B_c$ (diamonds) as a function of electron
density. The dashed lines are linear fits which extrapolate to the
critical electron density for the metal-insulator
transition.\vspace{0.20in}
}
\label{slope}
}
}
density data for $1/A$ are approximated well by a
linear dependence which extrapolates to the critical electron density
$n_c$ in a similar way to the behavior of the polarization field
$B_c$ (note that the tendency of the polarization field to vanish
near $n_c$ was also reported in Ref.~\cite{rem2}). We emphasize that
it has been verified with the help of the weak-field low-temperature
Hall effect measurements that the density of the delocalized
electrons in the metallic phase is practically coincident with $n_s$.

As has already been mentioned, the coefficient $A$ in the
linear-in-$T$ correction to conductivity of Eq.~(\ref{sigma}) is
determined by the Fermi liquid parameters \cite{aleiner}:

\begin{equation}
A=-\frac{(1+\alpha F_0^a)gm}{\pi\hbar^2n_s}. \label{A}\end{equation}
The factor $\alpha$ is equal to 8 in our case where the temperature
is small compared to the valley splitting \cite{private}. This
theoretical relation allows us to determine the many-body enhanced
$g$ factor and mass $m$ separately using the data for the slope $A$
and the product $gm$ as a function of $n_s$ (the latter is known from
independent measurements similar to those described in
Ref.~\cite{ferro}).

In Fig.~\ref{gm}, we show the so-determined values $g/g_0$ and
$m/m_b$ as a function of the electron density (the band mass $m_b$ is
equal to $0.19m_e$ where $m_e$ is the free electron mass). Note that
in the range of $n_s$ studied here, the low-temperature conductivity
$\sigma>8e^2/h$. The behavior of $g$ and $m$ at electron densities
below $n_s=3\times 10^{11}$~cm$^{-2}$ (corresponding to $r_s\approx
4.8$) turns out to be very different from that at electron densities
above this value. In the high $n_s$ region (lower $r_s$), the
enhancement of both $g$ and $m$ is relatively small, both values
slightly increasing with decreasing electron density in agreement
with earlier
\vbox{
\vspace{-16mm}
\hbox{
\hspace{2mm}
\psfig{file=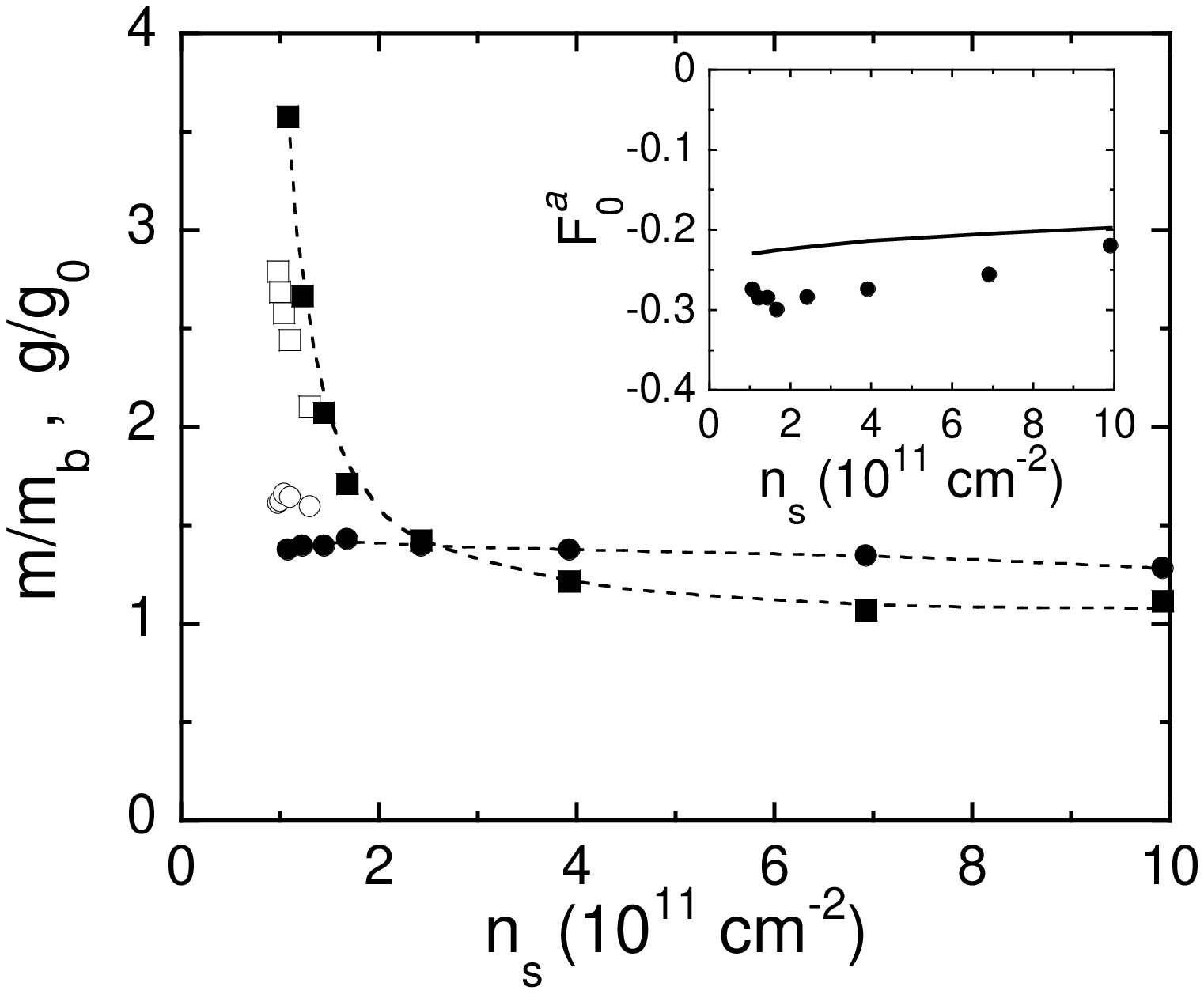,width=2.9in,bbllx=.5in,bblly=1.25in,bburx=7.25in,bbury=9.5in,angle=0}
}
\vspace{-0.4in}
\hbox{
\hspace{-0.15in}
\refstepcounter{figure}
\parbox[b]{3.4in}{\baselineskip=12pt \egtrm FIG.~\thefigure.
Renormalization of the effective mass (filled squares) and $g$ factor
(dots) as a function of electron density. The data from
Ref.~\protect\cite{PGK} are shown by open symbols. The dashed lines
are guides to the eye. The inset compares the theoretical dependence
of the renormalization parameter $F_0^a$ on $n_s$ (solid line) with
the data (dots) calculated using Eq.~(\ref{F}) from our $g$
values.\vspace{0.20in}
}
\label{gm}
}
}
data \cite{afs}. Also, the renormalization of the $g$
factor is dominant compared to that of the effective mass, which is
consistent with theoretical studies \cite{renorm}. The dependence
$g(n_s)$ is described reasonably well by the theory: the inset of
Fig.~\ref{gm} compares the theoretical renormalization parameter
$F_0^a=-r_s/2(2r_s+\sqrt2)$ \cite{aleiner} to that calculated using
Eq.~(\ref{F}) and the data for $g(n_s)$.

In contrast, the renormalization in the low $n_s$ (critical) region,
where $r_s\gg 1$, is much more striking. As the electron density is
decreased, Fig.~\ref{gm} shows that the renormalization of the
effective mass overshoots abruptly while that of the $g$ factor
remains relatively small, $g\approx g_0$, without tending to
increase. Hence, the current analysis indicates that it is the
effective mass that is responsible for the drastically enhanced $gm$
value near the metal-insulator transition.

The present results for the effective mass and $g$ factor in the
critical region can be compared to the data of Ref.~\cite{PGK}
obtained by analysis of the Shubnikov-de Haas oscillations in
high-mobility Si MOSFETs. As seen from Fig.~\ref{gm}, data obtained
from these two different methods are similar. Thus, the
Fermi-liquid-based theory \cite{aleiner} is adequate in describing
the properties of dilute 2D electron systems.

It is important to discuss another consequence of the theory
\cite{aleiner}: the slope $A$ of the temperature dependence of the
conductivity should increase as the ferromagnetic instability in a
dilute 2D electron system is approached. Since renormalization
parameters have not been theoretically calculated in the limit
$r_s\gg 1$, the simplest scenario of the ferromagnetic instability is
prompted by Eq.~(\ref{F}): $F_0^a\rightarrow -1$ causes the effective
$g$ factor (and the
\vbox{
\vspace{-16mm}
\hbox{
\hspace{2mm}
\psfig{file=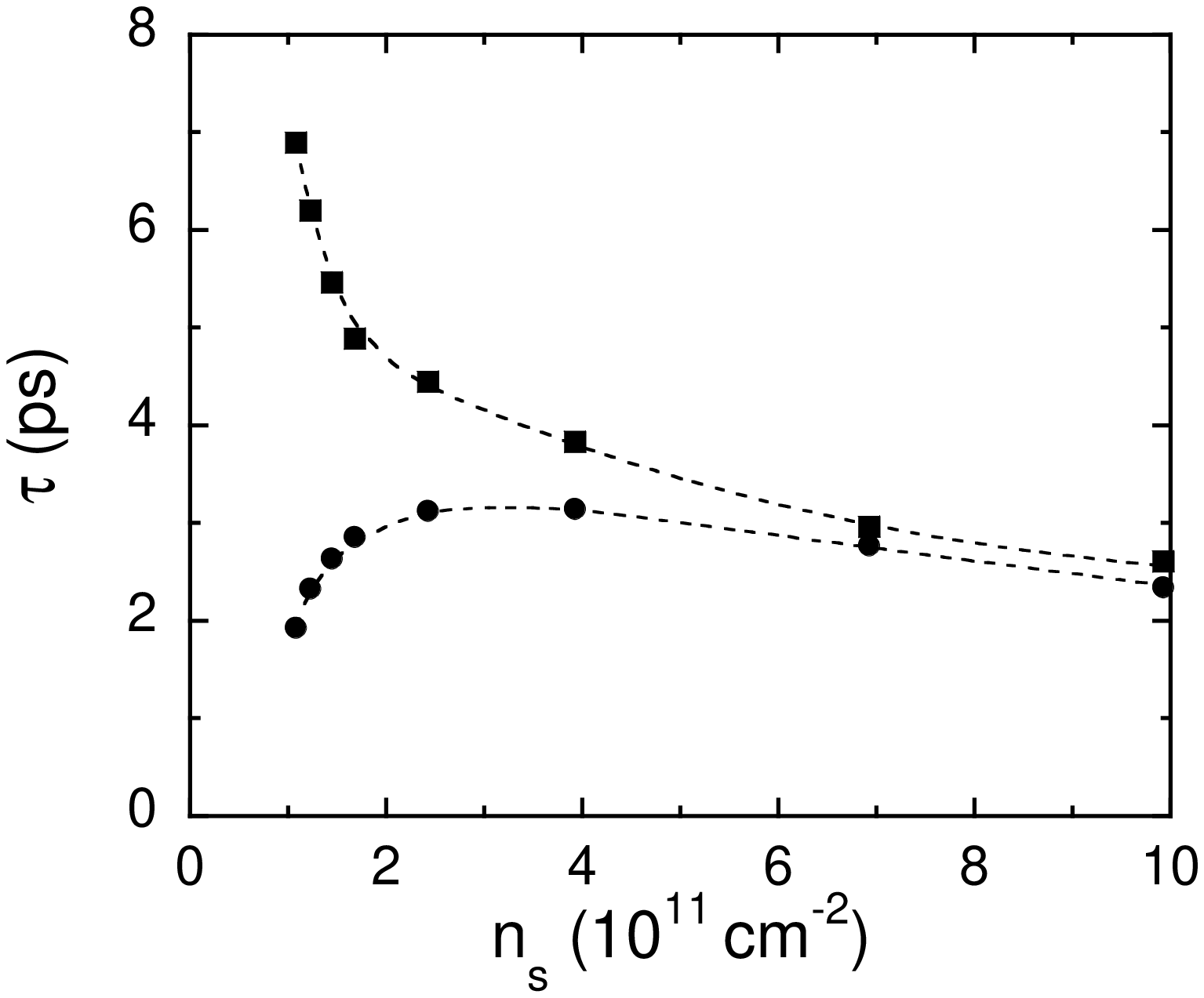,width=2.9in,bbllx=.5in,bblly=1.25in,bburx=7.25in,bbury=9.5in,angle=0}
}
\vspace{-0.4in}
\hbox{
\hspace{-0.15in}
\refstepcounter{figure}
\parbox[b]{3.4in}{\baselineskip=12pt \egtrm FIG.~\thefigure.
The elastic relaxation time versus electron density at a temperature
of 0.1~K, assuming $m=m_b$ (dots), and taking into account the
renormalization of $m$ (squares). The dashed lines are guides to the
eye.\vspace{0.20in}
}
\label{tau}
}
}
slope $A$) to diverge. Experimentally, the slope
$A$ tends to diverge near the critical electron density in a way
similar to the behavior of the product $gm$ seen in Fig.~\ref{slope}.
This is consistent with the conclusion of Ref.~\cite{ferro} about the
possibility of ferromagnetic instability in this electron system. At
the same time, the simplest scenario of a {\em diverging $g$ factor}
is not the case; instead, it is the growing {\em effective mass}
which controls the anomalous behavior of the dilute 2D electron
system near the metal-insulator transition.

The effective mass enhancement was traditionally considered to be
small, and, therefore, the value $m\approx m_b$ was used to calculate
some of the important system parameters, e.g., the elastic relaxation
time $\tau$ extracted from mobility. In Fig.~\ref{tau}, we compare
the so-determined $\tau$ (circles) with that calculated taking into
account the enhancement of the effective mass (squares). As seen from
the figure, in the range of electron densities studied, the corrected
$\tau$ keeps increasing down to the lowest $n_s$. Therefore, the
mobility drop at low $n_s$ in high-mobility Si MOSFETs (see, e.g.,
Ref.~\cite{kravchenko94}) turns out to come from the $m$ enhancement
rather than from the decrease in $\tau$, although the value $\tau$ is
still expected to vanish in the insulating phase. The observed
behavior of $\tau$ is consistent with that of the temperature range
corresponding to the ballistic regime (see Fig.~\ref{fan}), which
gives additional confidence in our analysis down to the vicinity of
the metal-insulator transition. Finally, values of $\tau$ much larger
than those previously estimated yield appreciably smaller quantum
level widths in perpendicular magnetic fields, which helps to
understand why the Shubnikov-de~Haas oscillations survive near the
metal-insulator transition, as well as the origin of the oscillations
of the metal-insulator phase boundary as a function of
(perpendicular) magnetic field \cite{phb}.

In summary, we have studied the temperature-dependent conductivity in
a wide range of electron densities above the critical electron
density for the metal-insulator transition. Using the recent theory
of interaction-driven corrections to conductivity \cite{aleiner}, we
extract Fermi-liquid parameters from the experimental data and
determine the many-body enhanced $g$ factor and the effective mass.
The tendency of the slope $A$ of the temperature dependence of the
conductivity to diverge near the critical density is in agreement
with the suggested ferromagnetic instability in this electron system
\cite{ferro}. Unexpectedly, it is found to originate from the growing
effective mass rather than the $g$ factor. In addition, the mass
enhancement is found to be responsible for the previously
underestimated values of elastic scattering time near the
metal-insulator transition.

We gratefully acknowledge discussions with I.~L. Aleiner, P.~T.
Coleridge, L.~I. Glazman, D. Heiman, J.~P. Kotthaus, B.~N. Narozhny,
and A. Punnoose. This work was supported by NSF grants DMR-9803440
and DMR-9988283, RFBR grant 01-02-16424, Forschungspreis of A. von
Humboldt Foundation, and the Sloan Foundation. TMK acknowledges
support through NSF grant PHY99-07949.





\end{multicols}
\end{document}